# Two-component energy spectrum of cuprates in the pseudogap phase and its evolution with temperature and at charge ordering

Lev P. Gor'kov[1,2] and Gregory B. Teitel'baum[3]*

In the search for mechanisms of high-temperature superconductivity it is critical to know the electronic spectrum in the pseudogap phase from which superconductivity evolves. The lack of angle-resolved photoemission data for every cuprate family precludes an agreement as to its structure, doping and temperature dependence and the role of charge ordering. Here we show that, in the entire Fermi-liquid-like regime that is ubiquitous in underdoped cuprates, the spectrum consists of holes on the Fermi arcs and an electronic pocket. We argue that experiments on the Hall coefficient identify the latter as a permanent feature at doped hole concentration $x$>0.08-0.10, in contrast to the idea of the Fermi surface reconstruction via charge ordering. The longstanding issue of the origin of the negative Hall coefficient in YBCO and Hg1201 at low temperature is resolved: the electronic contribution prevails as mobility of the latter (evaluated by the Dingle temperature) becomes temperature independent, while the mobility of holes scattered by the short-wavelength charge density waves decreases.

---

[1]NHMFL, Florida State University, 1800 East Paul Dirac Drive, Tallahassee Florida 32310, USA; [2]L.D. Landau Institute for Theoretical Physics of the RAS, Chernogolovka 142432, Russia; [3]E.K. Zavoiskii Institute for Technical Physics of the RAS, Kazan 420029, Russia.  *e-mail: grteit@kfti.knc.ru.

**Introduction**

Commonly, the transition into the superconducting state occurs in metals at a critical temperature $T_c$ from a normal phase that can be characterized by a well defined Fermi surface (FS). The high-transition-temperature (HT$_c$) cuprates dramatically deviate from the properties of the ordinary Fermi liquid (FL) in metals in that the pseudogap (PG) phase precedes the onset of superconductivity. ARPES reveals coherent excitations only within the so-called "Fermi arcs" [1, 2] (FAs). Another spectrum branch - small electronic pocket - manifests itself in quantum oscillations [3-5] (QOs). But ARPES is unavailable as yet for every material, while QOs are observable only at low temperatures. Therefore the nature of the pocket and its very existence at other temperatures has been debated for a long time [3-7]. With the tendency to a charge order (CO) transition revealed in the recent X-rays experiments [8-11], the view currently prevailing in the literature is that the pocket appears as the result of Fermi surface reconstruction [6, 7, 12-15] at a CO transition. We show that at doped hole concentrations $x>0.08-0.10$ the experimental Hall coefficient identifies the pocket as a *permanent* feature, in contrast to the idea of FS reconstruction at the charge ordering phase transition.

Recently it was found [16] that below a temperature $T^{**}(x)$ ($T_S<T<T^{**}(x)<T^*(x)$ with $T^*$ and $T_S$ being the PG and superconducting transition temperatures respectively) in the phase diagram of underdoped (UD) cuprates there exists a broad region in temperature and doping level in which resistivity displays a quadratic temperature dependence similar to that in a Fermi liquid. We prove that such a Fermi liquid-like charge transport regime is an ubiquitous feature of the pseudogap phase of UD YBCO, LSCO and Hg1201 in which the Fermi arc carriers play the important role. This opens a new avenue for studying the energy spectrum of excitations in cuprates. Below we determine the *bulk microscopic* characteristics of UD cuprates from the experimental data of resistivity and Hall coefficient. (The Fermi arcs in YBCO and Hg1201 were independently detected in the recent ARPES experiments [17, 18]).

In that follows, we initially consider the limit of low dopant concentration; as we demonstrate, in this regime the coherent excitations on the Fermi arcs (FAs) represent the only branch (hole-like) of the excitation spectrum in the conduction network of UD cuprates.

Among the three UD cuprates, Hg1201, YBCO and LCSO the most detailed information is available for LSCO [19, 20]; these data were used in most plots below. For the two other UD cuprates, Hg1201and YBCO, experimental data are scattered over the literature and results vary from not being obtained on one and the same crystalline sample. Low field data on the transverse magneto-resistance were recently published for Hg1201 [21].

In general, we find the good agreement between results from all three the cuprate families supporting thereby the idea that their unordinary properties originate from one and the same structural element- the $CuO_2$ plane.

As mentioned above, there is no consensus as yet between different groups concerning the origin of small pockets (and even of their number). (Note in passing, however, that the specific heat data at low temperatures [4] give support to only one electron pocket). The outstanding question is whether the pocket(s) is a mere band feature or is formed at a FS reconstruction in the process of a hypothetic phase transition [6, 7, 12-15]. Therefore the above analysis was repeated in more details at higher concentration, this time assuming for the system a two-component spectrum comprised of "holes" on the Fermi arcs and a small electron pocket coexisting on equal footing. It turned out that the electron pocket manifests itself for the first time at hole concentrations $p \geq 0.08 - 0.10$ and then becomes present in the PG phase *at all* temperatures.

It is shown below that this result is consistent with the temperature dependence of the Hall coefficient in YBCO and Hg1201 [13, 14] at low temperatures (and in the high magnetic field). The role of electron pocket is further discussed in the light of recent experiments revealing the onset of a charge density order at low temperatures [8-12].

**Results**

**Fermi-liquid-like resistivity regime.** Quadratic $T$-dependences for the resistivity and $\cot(\theta_H)$ (here $\theta_H$ is the so-called Hall angle) were known for a long time but could not be consistently explained in terms of a large Fermi surface [22] (see Supplementary information **SI**1). The analysis below unambiguously relates the FL-like regime with the carriers (holes) on FAs.

The theoretical model for the FA carriers was introduced in [23]. The parameters in the theoretical expressions for the resistivity $\rho(T) = [m*/n_{eff}e^2\bar{\tau}_{FA}(T)]$ and for the Hall coefficient $R_H = 1/|e|cn_{eff} > 0$ are defined as follows.

The effective number of carriers $n_{eff}$ is $n_{eff} = (\Delta\varphi p_F^2 / \pi^2)(s/c)$. At low hole concentrations and for the isotropic "bare" energy spectrum $\Delta\varphi p_F$ is the arc's length. Here $c$ is the lattice constant in the direction perpendicular to the $CuO_2$-plane; $s$ -the number of conducting layers (see Supplementary information **SI**2).

In the scattering rate $1/\bar{\tau}_{FA}(T) = 1/\tau_{h,imp} + 1/\tau_{FA}(T) \equiv 1/\tau_h$ the first term stands for scattering on defects; $1/\tau_{FA}(T) = 8\pi|V(1;2)|^2 T^2 / 3\hbar\Delta\varphi\varepsilon_F \times \sqrt{2[(K/4p_F)-1]}$ accounts for the inelastic Umklapp processes, $V(1;2) \sim 1$ is a dimensionless matrix element for the short-range electron-electron interactions; $m* = Z^{-1}m$ is the renormalized band mass, $Z$ the residue factor at the pole of the Green function; $\varepsilon_F = p_F^2/2m^*$ is the Fermi energy; in $[(K/4p_F)-1]$ $\vec{K} = (2\pi/a, 2\pi/a)$ is the Umklapp vector (see [6] and Supplementary information **SI**2). *All* these parameters can be measured; in particular, the plots in Fig.1 used the data from [19, 20]. The factor $T^2$ in the theoretical expression for $1/\tau_{FA}(T)$ gives the experimentally observed $T^2$- resistivity (in the clean limit or at higher temperatures, see below).

The Fermi arcs at the nodal points in the energy spectrum in UD cuprates are shown schematically in Fig.1a (insert); experimentally the Fermi arcs are known to be not far from the four nodal points $(\pm\pi/2, \pm\pi/2)$ and centered on the two diagonals of the tetragonal BZ.

With the mechanism underlying the anti-nodal gaps being still far from understood, the notion of the FAs remains a *phenomenological concept*. In particular, it suggests that all the doped holes go to FAs, an assumption confirmed by proportionality of the Fermi arc length to the dopant concentration $x$ (Fig. 1a,b).

In Fig. 1b the value of $\Delta\varphi/|V(1;2)|$ at small $x$ is superimposed on $\Delta\varphi(x)$ derived from Hall coefficient data $R_H(x,T) > 0$ [19, 20]. The procedure allowed determining the matrix element $|V(1,2)|$ =0.63. The value $(K/4p_F)-1$ is known experimentally [2]. The square root $\sqrt{2[(K/4p_F)-1]}$ varies insignificantly between $1/4 \div 1/2$. Correspondingly, in the following $1/\tau_{FA}(T) = AT^2/\Delta\varphi\varepsilon_F$ with $A$ between $1.5 - 0.8$ can be used as a reasonable estimate at *all* concentrations. Substituting m* = $4m_0$ for the effective mass in LSCO [24] and $p_F \approx \pi/\sqrt{2}a$ one finds $\varepsilon_F$ = 3000K.

The values of $\Delta\varphi(x)$ deduced from the Hall measurements for YBCO [25] at first fall on the same line. Stars correspond to $\Delta\varphi(x)$ for Hg1201 ($x$=0.075 [16]). (Numbers for the effective density of carriers $n_{eff} = \Delta\varphi p_F^2 s/c\pi^2$ were obtained from $R_H(x,T_+)$ [20] taken at a temperature $T_+$ slightly above the temperature of superconducting transition $T_c(x)$).

The quadratic $T$-dependence of $1/\tau_{FA}(T)$ for Hg1201 in Fig. 1c was confirmed by our re-plotting the data [21] making use of the theoretical expression: $\delta\rho/\rho_0 = (\tau_{FA}\omega_c^*)^2$ for the magnetoresistivity (see Supplementary information **SI**2).

At higher temperatures the resistivity behavior may deviate from the $T^2$ dependence. Plotted in Fig. 1d is the experimental resistivity multiplied by the number of carriers calculated from the Hall coefficient data [20]; in the Boltzmann approach the product is proportional to the inverse relaxation time. One finds that the $T^2$ dependence is preserved even at temperatures higher than $T^{**}(x)$: the concept of FAs actually remains applicable even at temperatures close to the PG temperature $T^*(x)$ [22].

The inverse residual resistivity for LSCO is plotted in Fig. 2a. Note that, in view of the strong tendency in this material to localization at lower concentrations, the proportionality of this contribution to the total conductivity to the number of dopants' is by itself a non-trivial fact. (One can think of this characteristic as of "conductivity of the mobile charge carriers").

Thus the whole body of experimental data on the transport properties of UD cuprates at temperatures $T > 100K$ can be accounted for in terms of only one charge component, namely of the holes on FAs. There are however some experimental features at the *same* temperatures that do not allow fully excluding the existence of an electron pocket in the PG phase, as discussed in the next section.

**Concentration dependence of the Hall coefficient and the electron pocket.** Common to all three families of cuprates is some irregular dependence on the hole concentration of several characteristics at the two compositions: $x_{1c} \approx 0.08$ and $x_{2c} \approx 0.12$. Compare, for instance, the *x*-dependence e.g. of $H_{c2}$ and $T_S$ [26] and the concentration dependence of the parameters shown in Figs. 1, 2. (See e. g. Fig. 2a for the inverse residual resistivity). From the (*T*, *x*)-phase diagram in Fig. 2b one sees that the interval of the FL-like regime in UD cuprates shrinks towards the same concentrations: $x_{1c} \approx 0.08$ and $x_{2c} \approx 0.12$. Regarding $x_{2c} \approx 0.12$, the consensus in the literature is [8, 9, 12] that at this concentration the tendency in UD cuprates to a charge ordering (CO) becomes important.

We argue that the deviations in Fig. 1b of $\Delta\varphi(x)$ from its initial linear *x* - dependence as seen in the Hall coefficient for LSCO together with other features at $x_{1c} \approx 0.08$ signify exactly the same physics as in the experiments [13] on QOs in UD YBCO, namely the first occurrence of the electron pocket in LSCO at $x_{1c} \approx 0.08$. (Why the linear in *x* dependence of $\Delta\varphi(x)$ obtained from the resistivity may continue to be seen at higher concentration will be discussed later).

**Activation temperature dependence of the Hall coefficient.** The Hall numbers for LSCO [19, 20] have been parametrized in [27, 28] using:

$$n_{Hall} = n_0(x) + n_1 \exp[-\Delta(x)/T]. \quad (1)$$

Here the numbers $n_{Hall}(x)$ and hence $n_0(x)$, $n_1$ are defined per unit cell, where $\Delta(x)$ is the activation energy. (Observe in the Supplementary information **SI3** how well such a decomposition satisfies the data [20]). Eq. (1) unambiguously defines the term $n_0(x)$ at the temperatures *much higher* than the CO temperature $T_{co}$. Deviations $n_0(x)$ from the expected linearity $n_0(x) = x$ become noticeable in the Hall coefficient *already at room temperatures* (see in Fig. 3b $R_H(T)$ plotted for $x = 0.12$).

The apparent similarity in Fig. 3a between these *experimental* features in $n_0(x)$ for LSCO and YBCO at $x > 0.08 - 0.1$ calls to mind the *low temperatures* experiments [13] in which the Hall coefficient in YBCO for the first time becomes negative, remarkably at the *same concentration* $x \approx 0.08$. (At 30-50K the Hall Effect is defined in magnetic fields strong enough to destroy superconductivity [13]).

**Hall coefficient for the interacting electrons and holes**. For a system with *only one* type of carrier such increase of $n_0(x)$ in Fig. 3a following from the positive Hall coefficient $R_H(T,x) > 0$ may signify nothing but an increase in the number of holes. That is not so if holes on FAs were interacting with a *small* pocket of electrons (see Supplementary information **SI2**) [29]).

Notably, the results for resistivity and the Hall coefficient would depend on the position of the pocket in the BZ. In most publications [6, 7, 12-14] devoted to the reconstruction of the FS in some hypothetical phase transition the pocket is presumed to form in a vicinity of FAs and, hence, would contribute to the resistivity via electron-electron umklapp processes.

Superconductivity according to recent X-ray experiments suppresses the charge ordering [8], thereby destroying such a pocket below $T_S$, the temperature of the superconducting transition. As mentioned above, this would be in contradiction to the low temperature specific heat data for *ortho-II* YBCO6.54 [4] that electrons in the pocket in the superconducting phase remain in the *normal state* down to 5K. (As shown in [30], for that the pocket must lie away from the Fermi arcs and at a position in the BZ of the high symmetry. This suggests one small electronic pocket at the $\Gamma$-point. (Unfortunately ARPES data along cuts crossing the $\Gamma$-point suffer from strong suppression of the photoemission intensity at the center of the BZ due to matrix element effects and can be hugely distorted, complicating the correct determination of the real dispersion [31]).

Parameters of the pocket are known from the frequency of QOs. The Fermi momentum is small $p_{Fe} \ll p_F$, and in the clean limit (i.e. at higher temperatures at which the role of defects becomes negligible) electrons cannot transfer their momentum to the lattice but have to follow carriers on FAs. That is, interactions bind holes and electrons together into a complex system. The corresponding parameter for relaxation between the two interacting sub-systems of electrons and holes $1/\tau_{eh}$ was estimated in [29]; $1/\tau_{eh}$ is also proportional to $T^2$ and of order of magnitude close to $1/\tau_{FA}(T)$.

With the number of electrons $n_e = (p_{Fe}^2/2\pi)(s/c)$ small compared to $n_{eff} = (\Delta\varphi p_F^2/\pi^2)(s/c)$ the theoretical expressions for the conductivity $\sigma$ and the Hall coefficient, $R_H = (ec\tilde{n}_{Hall})^{-1}$ become simpler (see Supplementary information **SI**2). In the limit $n_e \ll n_{eff}$,

$$\sigma = (e^2 n_{eff} \tau_{FA} / m^*)\{1 + \kappa(n_e / n_{eff})[1 - m_e / m^*]\} \quad (2)$$

and

$$\tilde{n}_{Hall} \approx \left[ n_{eff} - n_e(1 - \kappa - \kappa^2) \right] \quad (3)$$

Here $\kappa = (\tau_{eh} m^* / \tau_{FA} m_e)$. The correction to the Hall numbers $\tilde{n}_{Hall}$ in (3) would become *positive* at $\kappa > \kappa_c = (\sqrt{5} - 1)/2 \approx 0.62$, simulating thereby the seeming "increase" in number of holes in Fig. 3a.

From data [5] on QOs at $x \approx 0.1$ one obtains for YBCO $n_e / S_{BZ} \approx 0.036$ ($m_e/m^*$=0.5, $S_{BZ}$ is the area of BZ). Correspondingly, the electronic fraction in $\tilde{n}_{Hall}$ (3) is:
$n_{0e} \approx (n_e / S_{BZ})[(\kappa + 1/2)^2 - 5/4]$. Taking the value $\approx 0.02$ for the characteristic deviation in Fig.3a leads to the estimate for $\kappa = 0.84$. The relative contribution of the pocket into conductivity Eq. (2) turns out to be rather small ($\sim 0.15$) and not noticeable within the experimental accuracy [19,

20]. That agrees with the observed linear dependence of $\Delta\varphi(x)$ on $x$ in Fig.1a,b at concentrations $x > 0.08$).

The arguments above show the electronic pocket as a ubiquitous feature of the energy spectrum of UD cuprates in the PG phase above $x_{th} = 0.08 - 0.10$. Note that the existence of such threshold in the doping dependence suggests also possible structural changes in the system at these concentrations that may occur at inserting foreign atoms into the inter-plane "reservoir" blocks.

**Tendency to the charge density wave order.** The electron pocket having been established as a robust feature at least in such cuprates as LSCO and YBCO, there is a further inconsistency concerning its origin due to the FS reconstruction as suggested in [6, 7, 12-14]. Indeed, it is unclear firstly *when* in particular such reconstruction would take place: at $T_{co,onset}$, the onset temperature of the CO *fluctuations*, or at $T_{CO}$, the CO transition temperature. In fact, a quasi-static CDW order for YBCO6.54 sets in at $T_{CO} \approx 55K$, while $T_{co,onset} \approx 155K$ [10-12, 32]. Further, the Hall coefficients in YBCO and Hg1201 change sign at $\approx 25-30K$ and $\approx 15K$ respectively, starting to decrease already at $T \approx 100\,K$ [14].

In our discussion of the Hall coefficient in the CO phase below, we assume the pocket to already exist at temperatures much higher $T_{CO} \approx 55K$.

We now draw attention to the fact that the mobility of electrons in the CO phase is actually known *experimentally*. From the value of the Dingle temperature in *ortho II* YBCO, $T_D = 6K$, for the pocket [4] it follows that $1/\tau_e = 2\pi T_D = 30K$. Value of $1/\tau_{eh} \approx 1/\tau_{FA}(T) \approx T^2/\Delta\varphi\varepsilon_F$ at temperatures below $\approx 50K$ is less then $1/\tau_e$ ($\Delta\varphi \leq 0.1$, see Fig.(1 a, b)). The two sub-systems thus are decoupled already at $T \sim T_{CO} \approx 55K$. The Hall coefficient acquires the familiar form for the system of the two *independent* carriers:

$$R_H = \frac{1}{ec} \times \frac{n_{FA}(\bar{\tau}_{FA}/m^*)^2 - n_e(\tau_{e,imp}/m_e)^2}{[n_{FA}(\bar{\tau}_{FA}/m^*) + n_e(\tau_{e,imp}/m_e)]^2} \quad (4)$$

Thus, with $1/\tau_{e,imp} \sim 30-36K$ in Eq. (4) any further drop of $R_H(T) < 0$ with the temperature decrease can only come from a decrease of the mobility $\bar{\tau}_{FA}(T)/m^*$ of the positive carriers (holes on FAs). ($R_H(T_0) = 0$ defines the temperature $T_0$ [13]).

The inverse relaxation time for holes $1/\bar{\tau}_{FA}(T)$ increases as carriers on the Fermi arcs scatter on growing fluctuations of the incommensurate (IC) charge density wave (CDW). The CDW vector $Q \sim p_F \approx (\pi/\sqrt{2})a^{-1}$ having atomic size [8-11], the momentum transfer on scattering of the FAs excitations on a CDW with a short wavelength is large. (For electrons for which $p_{Fe} \ll Q \sim p_F$, such a rapidly oscillating CDW potential averages out, but all relevant contributions into the inverse scattering time are already included in the experimentally known value $1/\tau_{e,imp} \approx 30K$).

To be specific, let $M(\vec{p}; p_x \pm Q_x) = U_0 n(T) \exp(\pm Q_x)$ be the matrix element for scattering of a hole on the FAs with a momentum $\vec{p}$ by the CDW with parameter $n(T)$ (assuming uni-axial CDW). The inverse relaxation rate for holes scattering by CDW fluctuations $1/\tau_{FA,CDW}$ was calculated in [33]:

$$\frac{1}{\tau_{FA,CDW}} = (\pi^{3/2}/\sqrt{2})\bar{n}(T)^4 \left(\frac{\xi(T) p_F}{\Delta\varphi}\right)^2 \frac{1}{2}\sum_{\pm}\frac{1}{\varepsilon_F}\left[\frac{|U_0|^2}{\bar{E}(p_x \pm Q; p_y)}\right]^2. \quad (5)$$

In Eq. (5) $n(T)$ is in dimensionless units $n(T) \to \bar{n}(T)/a^2$, $U_0$ is a typical energy scale for cuprates (~ 1 eV).

In the expression (5) for the relaxation rate the coherence length $\xi(T)$ is large, $\xi \gg a$, so that the smallness of $\bar{n}(T)^4$ ($\bar{n}(T)$~0.1) is compensated by the factor $(\xi p_F / \Delta\varphi) \sim 10^3$ at $\xi p_F \sim 10^2$, $\Delta\varphi \sim 0.1$. As all characteristic energies for cuprates in (5) are $\sim 1 eV$, the value of $1/\tau_{FA,CDW}$ is of the order of a few hundred Kelvin, i.e. is much larger $1/\tau_e$. So up to the moment of the CO phase transition (in YBCO6.54 at the temperature $T_{CO}$~50 K) the FA carriers do not contribute significantly to any transport characteristic leaving electrons below $T_{CO}$ as the only mobile carriers. Eq.(4) is applicable below $T_{CO}$ as well, because due to disorder the CO parameter remains short-ranged even in the CO phase (see e.g. in [34]).

**Discussion**

In brief, the logical steps in the above argument are as follows. As pointed out in Introduction, no consensus yet exists concerning many of the enigmatic properties of the pseudogap phase (PG) which directly concerns the spectrum of the elementary excitations in UD cuprates. ARPES should allow immediate access to the PG energy spectrum, but in real materials the technique suffers from serious limitations. Among the most remarkable ARPES results is the observation of coherent excitations only on *disconnected* parts of the Fermi surface. The latter are known as Fermi arcs [1, 17, 18]. If a fundamental feature, the Fermi arcs should be present in all cuprates. To confirm this expectation, the supposition is made that the quadratic dependence of resistivity on temperature in the PG phase defines the unique Fermi liquid-like regime in UD cuprates. The regime is shown to follow from the presence of the Fermi arcs in the energy spectrum of the system and is an important feature common to all UD cuprates in the PG phase.

The basic *microscopic* parameters of the PG phase are determined from properties such as the conductivity and the Hall coefficient. In particular, the entire body of experimental data on transport properties of UD cuprates (with $x$< 0.08) in Figs.1, 2 at temperatures $T > 100K$ are well accounted for in terms of mobile holes on the Fermi arcs, confirming thereby their role as the only charged excitations participating in the kinetics. At higher dopant concentration our analysis shows that the energy spectrum of UD cuprates consists of holes on FAs and of a single electron pocket at the Γ-point,

This second branch of the spectrum, the electronic pocket, was first discovered in 2007 via observation of quantum oscillations (QOs) in the Hall coefficient [3]. QOs are studied in high magnetic fields and experimentally can be observed *only at low* enough temperatures. Therefore it became common in the literature to view this branch in the spectrum of UD cuprates as a low-temperature feature. In this scenario the electron pocket is predicted to appear due to the Fermi surface reconstruction at the temperature $T_{CO}$ of the CO phase transition (in YBCO~50K).

However this is not true: the experimental features in Fig. 3 at higher dopant concentrations demonstrate the presence of the electron pocket in the entire Fermi-liquid –like regime of the PG phase.

By itself, the FS reconstruction scenario does not specify when exactly such reconstruction must take place. According to [14], the negative contributions from the pocket start to decrease the Hall and Seebeck coefficients already at temperatures >100 K. Fig. 3 shows the signature of the electronic pocket at even higher temperatures (see Supplementary information **SI3**)**.**

Regarding the suggested reconstruction of the FS at the CO phase transition in [6, 7, 12-15], note in passing that a "diamond-shaped" pocket (see in [15]) could be feasible theoretically only for a charge density wave with a *two-directional* structural vector. As we argue above, the experimental behavior of the Hall Effect in the CO phase can be well-understood assuming a *uni-axial* charge density wave. The actual structure of the CO phase in YBCO is not yet resolved in X-ray experiments [9].

The residual metallic specific heat contribution from the pocket deep in the superconducting phase of YBCO [4] appears as the most decisive argument against the Fermi reconstruction scenario. Such behavior is not an isolated fact: the residual Sommerfeld coefficient is non-zero also in Hg1201, although there is a higher degree of uncertainty in its value (J. Kemper; private communication).

To summarize: The main results of the manuscript are as follows. At low doping level analysis of resistivity and the Hall coefficient identify the coherent excitations on the Fermi arcs as the only charge carriers (holes) in the system. At higher doping level deviation of the Hall numbers in LSCO and YBCO from proportionality to dopants concentration finds a natural explanation as due to the contribution from a small pocket of electrons *dragged* by holes on the Fermi arcs. On lowering temperature the Fermi arc carriers scatter strongly by fluctuations of incommensurable charge density waves, their mobility rapidly decreases and the contribution of holes to the transport properties gives way to that of electrons on the pocket.

**References**

[1] Yoshida, T. *et al*. Pseudogap, Superconducting Gap, and Fermi Arc in High-$T_c$ Cuprates Revealed by Angle-Resolved Photoemission Spectroscopy. *J. Phys. Soc. Jpn.* **81**, 011006 (2012).

[2] Yoshida, T. *et al*. Systematic doping evolution of the underlying Fermi surface of $La_{2-x}SrCuO_4$. *Phys. Rev. B* **74**, 224510 (2006).

[3] Doiron-Leyraud, N. *et al*. Quantum oscillations and the Fermi surface in an underdoped high-$T_c$ superconductor. *Nature* **447**, 565-568 (2007).

[4] Riggs, S. C. *et al*. Heat capacity through the magnetic-field-induced resistive transition in an underdoped high-temperature superconductor. *Nature Phys.* **7**, 332-335 (2011).

[5] Vignolle, B. *et al*. Quantum oscillations and the Fermi surface of high-temperature cuprate superconductors. *Comptes Rendus Physique*, **12**, Issue 5, 446-460 (2011).

[6] Taillefer, L. Fermi surface reconstruction in high-Tc superconductors. *J. Phys.: Condens. Matter* **21** 164212 (2009).

[7] Sebastian, S. E. *et al*. Normal-state nodal electronic structure in underdoped high-$T_c$ copper oxides. *Nature* **511**, 61–64 (2014).

[8] Chang, J. *et al*. Direct observation of competition between superconductivity and charge density wave order in $YBa_2Cu_3O_{6.67}$. *Nature Phys.* **8**, 871-876 (2012).

[9] Ghiringhelli, G. *et al*. Long-Range Incommensurate Charge Fluctuations in $(Y,Nd)Ba_2Cu_3O_{6+x}$. *Science* **337**, 821-825 (2012).

[10] Blackburn, E. *et al*. Inelastic x-ray study of phonon broadening and charge-density wave formation in ortho-II-ordered $YBa_2Cu_3O_{6.54}$. *Phys. Rev. B* **88**, 054506 (2013).

[11] Blackburn, E. *et al*. X-Ray Diffraction Observations of a Charge-Density-Wave Order in Superconducting Ortho-II $YBa_2Cu_3O_{6.54}$ Single Crystals in Zero Magnetic Field. *Phys. Rev. Lett.* **110**, 137004 (2013).

[12] Wu, T. *et al*. Magnetic-field-induced charge-stripe order in the high-temperature superconductor $YBa_2Cu_3O_y$. *Nature* **477**, 191-194 (2011).

[13] LeBoeuf, D. *et al*. Lifshitz critical point in the cuprate superconductor $YBa_2Cu_3O_y$ from high-field Hall effect measurements. *Phys. Rev. B* **83**, 054506 (2011).

[14] Doiron-Leyraud, N. *et al*. Hall, Seebeck, and Nernst Coefficients of Underdoped $HgBa_2CuO_{4+\delta}$: Fermi-Surface Reconstruction in an Archetypal Cuprate Superconductor. *Phys. Rev. X* **3**, 021019 (2013).

[15] Tabis, W. *et al*. Connection between charge-density-wave order and charge transport in the cuprate superconductors. Preprint at http://arXiv:1404.7658.

[16] Barišić, N. *et al*. Universal sheet resistance and revised phase diagram of the cuprate high-temperature superconductors. *Proc. Natl. Acad. Sci.* **110**, 12235 (2013).

[17] Hossain, M.A. *et al*. Controlling the self-doping of $YBa_2C_3O_{7-\delta}$ polar surfaces: From Fermi surface to nodal Fermi arcs by ARPES, *Nature Phys*. **4**, 527 (2008).

[18] Vishik I. M. *et al*., Angle-resolved photoemission spectroscopy study of $HgBa_2CuO_{4+\delta}$, *Phys. Rev. B* **89**, 195141 (2014)

[19] Ando, Y., Kurita, Y., Komiya, S., Ono, S. and Segawa, K. Evolution of the Hall Coefficient and the Peculiar Electronic Structure of the Cuprate Superconductors. *Phys. Rev. Lett*. **92**, 197001(2004).

[20] Ono, S., Komiya, S. and Ando, Y. Strong charge fluctuations manifested in the high-temperature Hall coefficient of high-$T_c$ cuprates. *Phys. Rev. B* **75**, 024515 (2007).

[21] Chan, M. K. *et al*. Validity of Kohler's rule in the pseudogap phase of the cuprate superconductors. Preprint at http://arXiv:1402.4472v1.

[22] Gor'kov, L. P. and Teitel'baum, G. B. Mobility and its temperature dependence in underdoped $La_{2-x}Sr_xCuO_4$ interpreted as viscous motion of charges. *Phys. Rev. B* **72**, 180511 (R) (2008).

[23] Gor'kov, L. P. Kinetics of excitations on the Fermi arcs in underdoped cuprates at low temperatures. *Phys. Rev. (Rapid Comm.) B* **88**, 041104 (2013).

[24] Padilla, W. J. *et al*. Constant effective mass across the phase diagram of high-$T_c$ cuprates, *Phys. Rev. B* **72**, 060511(R) (2005).

[25] Segawa, K., Ando, Y. Intrinsic Hall response of the $CuO_2$ planes in a chain-plane-composite system of $YBa_2Cu_3O_y$. *Phys. Rev. B* **69**, 104521 (2004).

[26] Grissonnanche, G. *et al*. Direct measurement of the upper critical field in a cuprate superconductor. *Nature Commun*. **5**, 3280 (2014).

[27] Gor'kov, L. P. and Teitel'baum, G. B. Interplay of Externally Doped and Thermally Activated Holes in $La_{2-x}Sr_xCuO_4$ and Their Impact on the Pseudogap Crossover. *Phys. Rev. Lett*. **97**, 247003 (2006).

[28] Gor'kov, L. P. and Teitel'baum, G. B. The two-component physics in cuprates in the real space and in the momentum representation. *J. of Phys.: Conf. Series* **108**, 012009 (2008).

[29] Gor'kov, L. P. and Teitel'baum, G. B. Two regimes in conductivity and the Hall coefficient of underdoped cuprates in strong magnetic fields, *J. Phys.: Condens. Matter* **26,** 042202 (2014).

[30] Gor'kov, L. P. Quantum oscillations in the vortex state of underdoped $YBa_2Cu_3O_{6.5}$ and other multi-band superconductors, *Phys. Rev. B* **86**, 060501 (2012).

[31] Inosov, D. S. *et al*. Momentum and Energy Dependence of the Anomalous High-Energy Dispersionin the Electronic Structure of High Temperature Superconductors, *Phys. Rev. Lett*. **99**, 237002 (2007)

[32] LeBoeuf, D. *et al*. Thermodynamic phase diagram of static charge order in underdoped $YBa_2Cu_3O_y$, *Nature Phys*. **9**, 79–83 (2013).

[33] Gor'kov, L. P., Breakdown of the Fermi arcs in underdoped cuprates by incommensurate charge density waves, *Pis'ma v ZhETF,* **100**, 447 (2014).

[34] Wu, T. *et al.* Short-range charge order reveals the role of disorder in the pseudogap state of high-$T_c$ superconductors, arXiv:1404.1617.

**The legends for the figures**

**Figure 1. Carriers on the Fermi arcs in the conduction network of UD cuprates from the experimental transport data:**

**a**. The arc length divided by the matrix element $\Delta\varphi/V(1;2)$ as function of doping (from the resistivity data [19, 20]). Proportionality of the Fermi arcs' length to the dopants' concentration $x$ implies that all doped holes go to FAs. Insert: The dashed circles show schematically the "bare "Fermi surface of UD cuprates; the thick "patches" represent the Fermi arcs. At the tetragonal symmetry the latter are centered not far from the four points $(\pm\pi/2, \pm\pi/2)$ in the Brillouin zone (BZ). The electronic pockets not shown. Arrows show Umklapp processes for scattering of the two holes between the different Fermi arcs that could contribute into resistivity. **b.** The value of $\Delta\varphi/|V(1,2)|$ from the resistivity at small $x$ superimposed on $\Delta\varphi(x)$ derived from the Hall coefficient $R_H(x,T) > 0$. The procedure allows determining the value of the renormalized electron coupling $|V(1;2)| \approx 0.63$. Doping dependence of $\Delta\varphi(x)$ for YBCO from the Hall measurements [25] is also shown (diamonds). Star corresponds to $\Delta\varphi(T)$ for Hg1201 at $x$=0.075 [14]. Note again proportionality of $\Delta\varphi(x)$ to the concentration at small $x$, this time obtained using the Hall data. (The line is guide for eyes). **c**. The $T^2$ dependence of the scattering rate for holes on the arc in the Hg1201 (re-plotted from magnetotransport data [21]). **d**. Resistivity multiplied by $n_{\text{Hall}}(T, x)$ plotted against $T^2$ for LSCO at several $x$ (deduced from [20]). (The product is proportional to the inverse relaxation time). The $T^2$ dependence is preserved at temperatures even higher than $T^{**}(x)$; that is, the FAs concept is applicable at temperatures close to the PG temperature $T^*(x)$ [22].

**Figure 2. Irregularities at the two concentrations $x = 0.08 \div 0.15$:**

**a.** Plotted is the inverse residual resistivity ("conductivity of mobile charges") for LSCO; its proportionality to the dopant concentration is a non-trivial result because of the well-known tendency to localization at lower concentrations in this material. **b**. The generic ($T$, $x$)-phase diagram. The FL-like regime in UD cuprates shrinks towards concentrations: $x_{1c} \approx 0.08$ and $x_{2c} \approx 0.12$. (To the experimental accuracy, $T^*(x)$, the PG temperature falls between the dashed- and the dot-dashed lines). Open symbols mark the upper boundary of the $\rho \propto T^2$ regime, $T^{**}(x)$. Correspondingly, the filled symbols stand for the lower boundary (all lines are guides for eye).

**Figure 3**. **The temperature independent term $n_0(x)$ in Eq. (1):**

**a**. Deviations from the dashed straight line $n_0 = x$ line for $x > 0.08$ in the effective carrier number $n_0(x)$ for LSCO and YBCO (according to the data of [13, 20, 25]). Unlike the case for LSCO [20], for YBCO the hole number at the nominal oxygen content comes from different groups [13] and [25]. **b**. Fitting of the Hall coefficient temperature dependence Eq. (1) to the experimental data [19, 20] for LSCO with doping level $x=0.12$. The continuous line corresponds to $n_0$ as the fitting parameter. The dashed line – to the Eq. (1) with $n_0$ fixed ($n_0 = x \equiv 0.12$). The pronounced divergence of these curves at lower temperatures, starting at 350 K, is due to emergence of the electronic contribution. (See text below Eq. (3)).

**Acknowledgements**  The work of L. P. G. was supported by the NHMFL through NSF Grant No. DMR-1157490, the State of Florida and the U.S. Department of Energy; that of G. B. T. by the Russian Academy of Sciences through Grants No. P 20 and No. OFN 03.

**Competing financial interests**  The authors declare no competing financial interests.

Figure 1

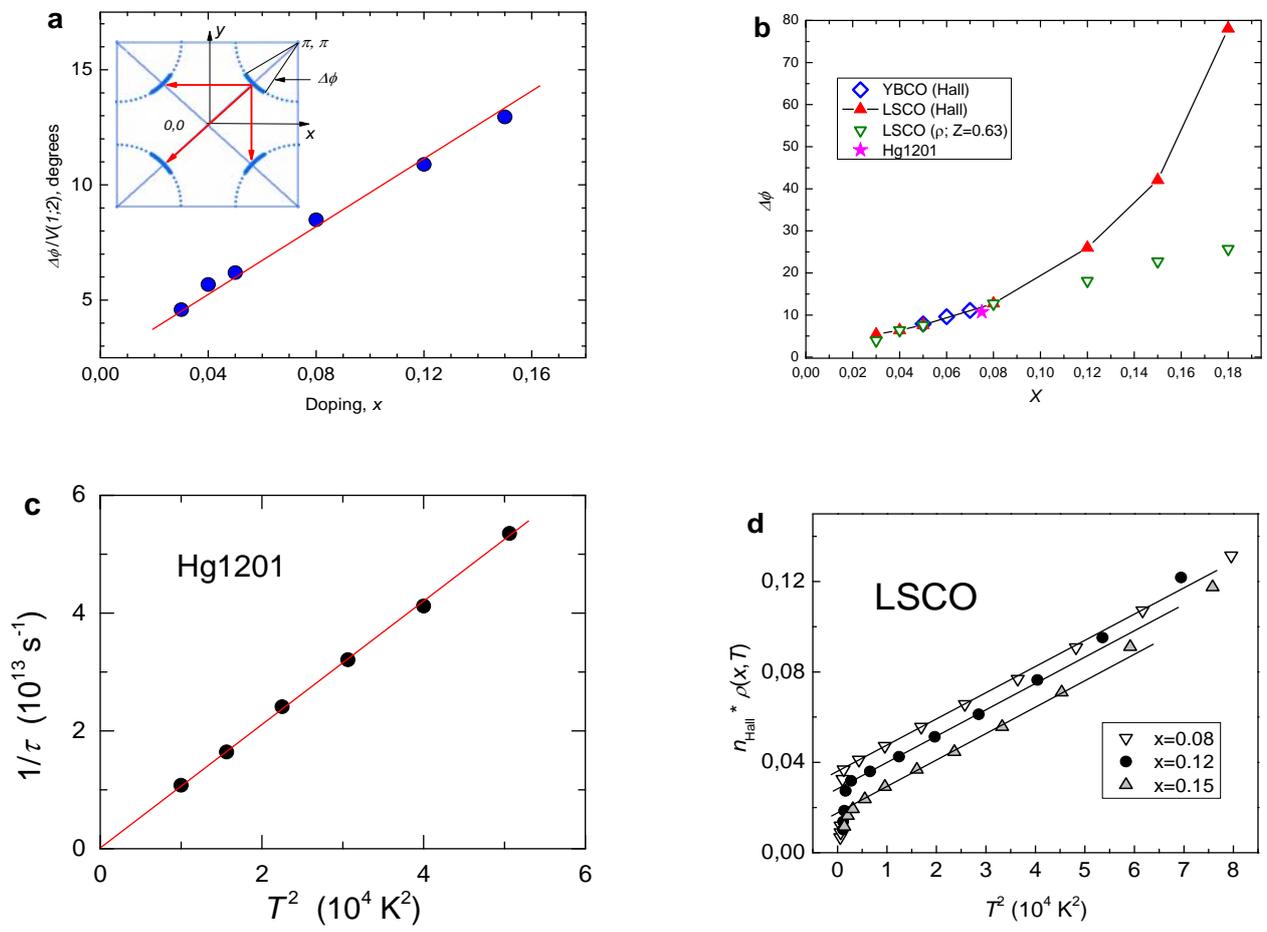

Figure 2

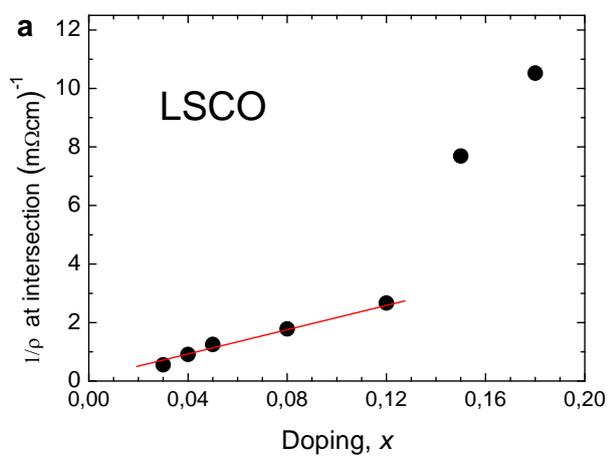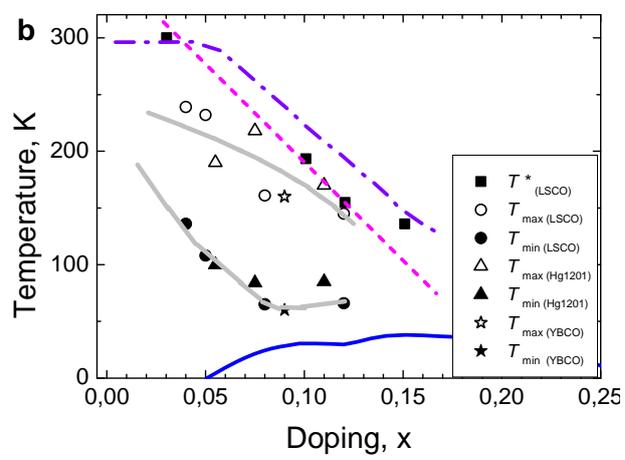

Figure 3

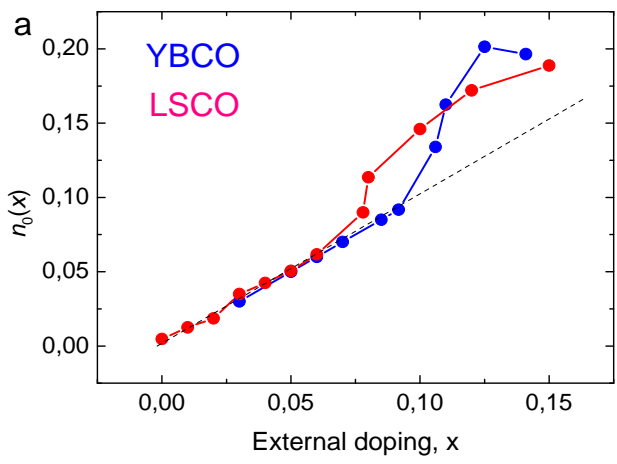 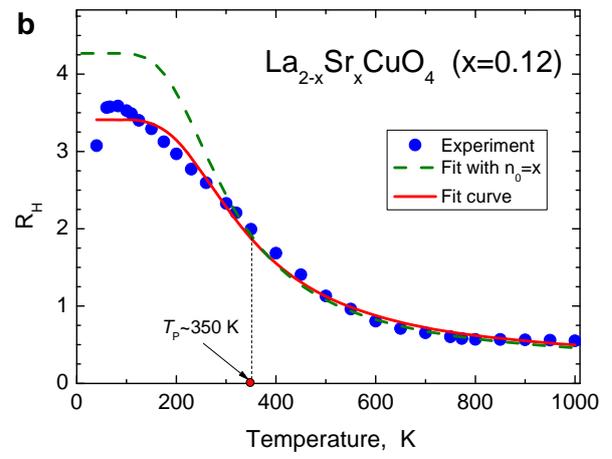